\documentclass[superscriptaddress,preprintnumbers,showpacs,amsmath,amssymb,prl,twocolumn,floatfix]{revtex4}

\usepackage{graphicx}
\usepackage{bm}
\usepackage{ifthen}
\usepackage{setspace}
\usepackage{color}

\def\ra{\rightarrow}
\newcommand{\GeV}{\ensuremath{\mathrm{Ge\kern -0.1em V}}}
\newcommand{\MeV}{\ensuremath{\mathrm{Me\kern -0.1em V}}}

\newcommand{\pp}{\ensuremath{p\overline{p}}}

\newcommand{\fb}{\ensuremath{\mathrm{fb}^{-1}}}
\newcommand{\bs}{\ensuremath{B_s^0}}
\newcommand{\bd}{\ensuremath{B^0}}
\newcommand{\bsd}{\ensuremath{B_s^0(B^0)}}
\newcommand{\bu}{\ensuremath{B^{+}}}

\newcommand{\mm}{\ensuremath{\mu^{+}\mu^{-}}}

\newcommand{\hh}{\ensuremath{h^{+}h^{\prime -}}}
\newcommand{\bsmm}{\ensuremath{\bs\ra\mm}}
\newcommand{\bdmm}{\ensuremath{\bd\ra\mm}}
\newcommand{\bsdmm}{\ensuremath{\bsd\ra\mm}}
\newcommand{\bjk}{\ensuremath{\bu\ra J/\psi~K^{+}}}
\newcommand{\bjpi}{\ensuremath{\bu\ra J/\psi~\pi^{+}}}
\newcommand{\bsjp}{\ensuremath{\bs\ra J/\psi~\phi}}

\newcommand{\brbsmm}{\ensuremath{\mathcal{B}(\bsmm)}}
\newcommand{\brbdmm}{\ensuremath{\mathcal{B}(\bdmm)}}
\newcommand{\Mmm}{\ensuremath{m_{\mu\mu}}}

\newcommand{\pting}{\ensuremath{\Delta\Omega}}
\newcommand{\iso}{\ensuremath{\mathit{I}}}
\newcommand{\nn}{\ensuremath{\nu_{N}}}
\newcommand{\ptmm}{\ensuremath{\vec{p}^{\:\mu\mu}_{T}}}
\newcommand{\pmm}{\ensuremath{\vec{p}^{\:\mu\mu}}}
\newcommand{\cdf}{CDF~II}

\newcommand{\dedx}{\ensuremath{dE/dx}}

\begin{document}


\title{

\bf{ Search for $\bsmm$ and $\bdmm$ Decays 
        with \cdf }} 



\begin{abstract}
 A search has been performed for \bsmm\ and \bdmm\ decays using $7\:\fb$ of integrated
 luminosity collected by the \cdf\ detector at the Fermilab Tevatron
 collider.  The observed number of \bd\ candidates is consistent with
 background-only expectations and yields an upper limit on the branching fraction of 
 $\brbdmm < 6.0 \times 10^{-9}$ at 95\% confidence level.  We observe an excess of
 \bs\ candidates.  The probability that the background processes alone could
 produce such an excess or larger is $0.27\%$.  The probability that the combination
 of background and the expected standard model rate of \bsmm\ could produce such an excess or larger
 is $1.9\%$. These data are used
 to determine $\brbsmm = (1.8^{+1.1}_{-0.9}) \times 10^{-8} $ and provide
 an upper limit of $\brbsmm < 4.0 \times 10^{-8}$ at 95\% confidence level.
\end{abstract}

\affiliation{Institute of Physics, Academia Sinica, Taipei, Taiwan 11529, Republic of China} 
\affiliation{Argonne National Laboratory, Argonne, Illinois 60439, USA} 
\affiliation{University of Athens, 157 71 Athens, Greece} 
\affiliation{Institut de Fisica d'Altes Energies, ICREA, Universitat Autonoma de Barcelona, E-08193, Bellaterra (Barcelona), Spain} 
\affiliation{Baylor University, Waco, Texas 76798, USA} 
\affiliation{Istituto Nazionale di Fisica Nucleare Bologna, $^{aa}$University of Bologna, I-40127 Bologna, Italy} 
\affiliation{University of California, Davis, Davis, California 95616, USA} 
\affiliation{University of California, Los Angeles, Los Angeles, California 90024, USA} 
\affiliation{Instituto de Fisica de Cantabria, CSIC-University of Cantabria, 39005 Santander, Spain} 
\affiliation{Carnegie Mellon University, Pittsburgh, Pennsylvania 15213, USA} 
\affiliation{Enrico Fermi Institute, University of Chicago, Chicago, Illinois 60637, USA}
\affiliation{Comenius University, 842 48 Bratislava, Slovakia; Institute of Experimental Physics, 040 01 Kosice, Slovakia} 
\affiliation{Joint Institute for Nuclear Research, RU-141980 Dubna, Russia} 
\affiliation{Duke University, Durham, North Carolina 27708, USA} 
\affiliation{Fermi National Accelerator Laboratory, Batavia, Illinois 60510, USA} 
\affiliation{University of Florida, Gainesville, Florida 32611, USA} 
\affiliation{Laboratori Nazionali di Frascati, Istituto Nazionale di Fisica Nucleare, I-00044 Frascati, Italy} 
\affiliation{University of Geneva, CH-1211 Geneva 4, Switzerland} 
\affiliation{Glasgow University, Glasgow G12 8QQ, United Kingdom} 
\affiliation{Harvard University, Cambridge, Massachusetts 02138, USA} 
\affiliation{Division of High Energy Physics, Department of Physics, University of Helsinki and Helsinki Institute of Physics, FIN-00014, Helsinki, Finland} 
\affiliation{University of Illinois, Urbana, Illinois 61801, USA} 
\affiliation{The Johns Hopkins University, Baltimore, Maryland 21218, USA} 
\affiliation{Institut f\"{u}r Experimentelle Kernphysik, Karlsruhe Institute of Technology, D-76131 Karlsruhe, Germany} 
\affiliation{Center for High Energy Physics: Kyungpook National University, Daegu 702-701, Korea; Seoul National University, Seoul 151-742, Korea; Sungkyunkwan University, Suwon 440-746, Korea; Korea Institute of Science and Technology Information, Daejeon 305-806, Korea; Chonnam National University, Gwangju 500-757, Korea; Chonbuk National University, Jeonju 561-756, Korea} 
\affiliation{Ernest Orlando Lawrence Berkeley National Laboratory, Berkeley, California 94720, USA} 
\affiliation{University of Liverpool, Liverpool L69 7ZE, United Kingdom} 
\affiliation{University College London, London WC1E 6BT, United Kingdom} 
\affiliation{Centro de Investigaciones Energeticas Medioambientales y Tecnologicas, E-28040 Madrid, Spain} 
\affiliation{Massachusetts Institute of Technology, Cambridge, Massachusetts 02139, USA} 
\affiliation{Institute of Particle Physics: McGill University, Montr\'{e}al, Qu\'{e}bec, Canada H3A~2T8; Simon Fraser University, Burnaby, British Columbia, Canada V5A~1S6; University of Toronto, Toronto, Ontario, Canada M5S~1A7; and TRIUMF, Vancouver, British Columbia, Canada V6T~2A3} 
\affiliation{University of Michigan, Ann Arbor, Michigan 48109, USA} 
\affiliation{Michigan State University, East Lansing, Michigan 48824, USA}
\affiliation{Institution for Theoretical and Experimental Physics, ITEP, Moscow 117259, Russia}
\affiliation{University of New Mexico, Albuquerque, New Mexico 87131, USA} 
\affiliation{Northwestern University, Evanston, Illinois 60208, USA} 
\affiliation{The Ohio State University, Columbus, Ohio 43210, USA} 
\affiliation{Okayama University, Okayama 700-8530, Japan} 
\affiliation{Osaka City University, Osaka 588, Japan} 
\affiliation{University of Oxford, Oxford OX1 3RH, United Kingdom} 
\affiliation{Istituto Nazionale di Fisica Nucleare, Sezione di Padova-Trento, $^{bb}$University of Padova, I-35131 Padova, Italy} 
\affiliation{LPNHE, Universite Pierre et Marie Curie/IN2P3-CNRS, UMR7585, Paris, F-75252 France} 
\affiliation{University of Pennsylvania, Philadelphia, Pennsylvania 19104, USA}
\affiliation{Istituto Nazionale di Fisica Nucleare Pisa, $^{cc}$University of Pisa, $^{dd}$University of Siena and $^{ee}$Scuola Normale Superiore, I-56127 Pisa, Italy} 
\affiliation{University of Pittsburgh, Pittsburgh, Pennsylvania 15260, USA} 
\affiliation{Purdue University, West Lafayette, Indiana 47907, USA} 
\affiliation{University of Rochester, Rochester, New York 14627, USA} 
\affiliation{The Rockefeller University, New York, New York 10065, USA} 
\affiliation{Istituto Nazionale di Fisica Nucleare, Sezione di Roma 1, $^{ff}$Sapienza Universit\`{a} di Roma, I-00185 Roma, Italy} 

\affiliation{Rutgers University, Piscataway, New Jersey 08855, USA} 
\affiliation{Texas A\&M University, College Station, Texas 77843, USA} 
\affiliation{Istituto Nazionale di Fisica Nucleare Trieste/Udine, I-34100 Trieste, $^{gg}$University of Udine, I-33100 Udine, Italy} 
\affiliation{University of Tsukuba, Tsukuba, Ibaraki 305, Japan} 
\affiliation{Tufts University, Medford, Massachusetts 02155, USA} 
\affiliation{University of Virginia, Charlottesville, Virginia 22906, USA}
\affiliation{Waseda University, Tokyo 169, Japan} 
\affiliation{Wayne State University, Detroit, Michigan 48201, USA} 
\affiliation{University of Wisconsin, Madison, Wisconsin 53706, USA} 
\affiliation{Yale University, New Haven, Connecticut 06520, USA} 
\author{T.~Aaltonen}
\affiliation{Division of High Energy Physics, Department of Physics, University of Helsinki and Helsinki Institute of Physics, FIN-00014, Helsinki, Finland}
\author{B.~\'{A}lvarez~Gonz\'{a}lez$^w$}
\affiliation{Instituto de Fisica de Cantabria, CSIC-University of Cantabria, 39005 Santander, Spain}
\author{S.~Amerio}
\affiliation{Istituto Nazionale di Fisica Nucleare, Sezione di Padova-Trento, $^{bb}$University of Padova, I-35131 Padova, Italy} 

\author{D.~Amidei}
\affiliation{University of Michigan, Ann Arbor, Michigan 48109, USA}
\author{A.~Anastassov}
\affiliation{Northwestern University, Evanston, Illinois 60208, USA}
\author{A.~Annovi}
\affiliation{Laboratori Nazionali di Frascati, Istituto Nazionale di Fisica Nucleare, I-00044 Frascati, Italy}
\author{J.~Antos}
\affiliation{Comenius University, 842 48 Bratislava, Slovakia; Institute of Experimental Physics, 040 01 Kosice, Slovakia}
\author{G.~Apollinari}
\affiliation{Fermi National Accelerator Laboratory, Batavia, Illinois 60510, USA}
\author{J.A.~Appel}
\affiliation{Fermi National Accelerator Laboratory, Batavia, Illinois 60510, USA}
\author{A.~Apresyan}
\affiliation{Purdue University, West Lafayette, Indiana 47907, USA}
% ...
\collaboration{CDF Collaboration\footnote{With visitors from $^a$Istituto Nazionale di Fisica Nucleare, Sezione di Cagliari, 09042 Monserrato (Cagliari), Italy,
$^b$University of CA Irvine, Irvine, CA  92697, USA,
$^c$University of CA Santa Barbara, Santa Barbara, CA 93106, USA,
$^d$University of CA Santa Cruz, Santa Cruz, CA  95064, USA,
$^e$CERN,CH-1211 Geneva, Switzerland,
$^f$Cornell University, Ithaca, NY  14853, USA, 
$^g$University of Cyprus, Nicosia CY-1678, Cyprus, 
$^h$Office of Science, U.S. Department of Energy, Washington, DC 20585, USA,
$^i$University College Dublin, Dublin 4, Ireland,
$^j$University of Fukui, Fukui City, Fukui Prefecture, Japan 910-0017,
$^k$Universidad Iberoamericana, Mexico D.F., Mexico,
$^l$Iowa State University, Ames, IA  50011, USA,
$^m$University of Iowa, Iowa City, IA  52242, USA,
$^n$Kinki University, Higashi-Osaka City, Japan 577-8502,
$^o$Kansas State University, Manhattan, KS 66506, USA,
$^p$University of Manchester, Manchester M13 9PL, United Kingdom,
$^q$Queen Mary, University of London, London, E1 4NS, United Kingdom,
$^r$University of Melbourne, Victoria 3010, Australia,
$^s$Muons, Inc., Batavia, IL 60510, USA,
$^t$Nagasaki Institute of Applied Science, Nagasaki, Japan, 
$^u$National Research Nuclear University, Moscow, Russia,
$^v$University of Notre Dame, Notre Dame, IN 46556, USA,
$^w$Universidad de Oviedo, E-33007 Oviedo, Spain, 
$^x$Texas Tech University, Lubbock, TX  79609, USA,
$^y$Universidad Tecnica Federico Santa Maria, 110v Valparaiso, Chile,
$^z$Yarmouk University, Irbid 211-63, Jordan,
$^{hh}$On leave from J.~Stefan Institute, Ljubljana, Slovenia, 
}}
\noaffiliation


\maketitle


Studies of flavor-changing neutral current (FCNC) decays have played an important role 
in formulating the theoretical description of particle physics known as 
the standard model (SM).  In the SM all neutral currents conserve flavor so that 
FCNC decays do not occur at lowest order. The decays of \bs\ mesons (with a quark content of 
$\overline{b}s$) and \bd\ mesons ($\overline{b}d$) into a dimuon pair (\mm )~\cite{qconj}
are examples of FCNC processes that can occur in the SM through higher order loop diagrams.
Their branching 
fractions are predicted in the SM to be $(3.2\pm0.2)\times10^{-9}$ and 
$(1.0\pm0.1)\times10^{-10}$, respectively~\cite{smbr}. A wide variety of beyond-SM 
theories predict significant increases over the SM branching fraction~\cite{tanb}, 
making the study of \bsmm\ and \bdmm\ decays  
one of the most sensitive indirect searches for new physics. Published upper 
limits~\cite{CDFPRL2008,D0PLB2010,LHCb} contribute significantly to our knowledge of 
the available new physics parameter space~\cite{rparity,dark,chi2,baek,buras}.

We report a search for \bsmm\ and \bdmm\ decays using \pp\ data corresponding to an integrated 
luminosity of $7\:\fb$ collected with the Collider Detector at Fermilab (\cdf).  
The sensitivity of this analysis is significantly improved with respect to the previous 
analysis~\cite{CDFPRL2008} due to the higher integrated luminosity of the event sample, a
20\% increase in the signal acceptance, and the use of an improved neural-network (NN) 
discriminant that provides approximately twice the background rejection for the 
same signal efficiency.   

A detailed description of the \cdf\ detector can be found in Ref.~\cite{Acosta:2004yw}. 
A charged particle tracking system provides precise vertex determination and momentum 
measurements in a pseudorapidity range $\left|\eta\right|<1.0$.
Additionally, the system measures the ionization per unit path length $\dedx$ for 
particle identification.  Beyond the
tracking detectors are electromagnetic and hadronic calorimeters, which are surrounded by
drift chambers used to detect muons in the central region (C) $\left|\eta\right|<0.6$ 
and the forward region (F) $0.6<\left|\eta\right|<1.0$.  

The online (trigger) requirements used to collect the data sample and the initial set of 
baseline requirements used in the analysis are the same as those described in Ref.~\cite{PRL1}.
The events are collected using a set of dimuon triggers~\cite{Acosta:2004yw} and  
must satisfy either of two sets of requirements
corresponding to different topologies: CC events have both muon candidates detected in the 
central region, while CF events have one central muon and another muon detected
in the forward region. Since the expected signal-to-background ratios are different, the 
two topologies are treated separately.  The acceptance of the analysis is improved by 20\% 
by using additional foward muon candidates and by using muon candidates that traverse detector 
regions previously excluded due to their rapidly
changing trigger efficiency.  The larger data sample has allowed us to obtain a detailed 
understanding of the trigger performance in these regions so that we can confidently include
these muon candidates in the current analysis.   The baseline selection requires high quality 
muon candidates with transverse momentum relative to the beam direction of $p_{T} > 2.0\:(2.2)\:\GeV/c$
in the central (forward) region.
The muon pairs are required to have an invariant mass in the range $4.669 < \Mmm < 5.969\:\GeV/c^2$
and are constrained to originate from a common well measured three-dimensional (3D) vertex.
A likelihood method~\cite{mixprl} together with a \dedx\ based selection~\cite{bhhprl} are used to
further suppress contributions from hadrons misidentified as muons.
The baseline requirements also demand that the measured proper decay length of the {\it B} candidate
$\lambda$ with its uncertainty $\sigma_\lambda$ satisfy
$\lambda/\sigma_\lambda > 2$;  the 3D opening angle between the momentum of the dimuon pair
and the displacement vector between the primary \pp\ collision vertex and the dimuon vertex
$\pting < 0.7~\rm{rad}$; and the {\it B}-candidate track isolation~\cite{isodef} $\iso > 0.50$.
There are $48\,279$ CC and $52\,179$ CF muon pairs that fulfill the trigger and baseline selection 
requirements.  

A sample of $\bjk$ events serves as a normalization mode. The $\bjk$ sample is collected using 
the same dimuon triggers and selection requirements so that common systematic uncertainties are
suppressed.  An additional requirement on the kaon candidate $p_T>1~\GeV/c$ is made to limit 
the $p_T$ range to a region where the tracking efficiency is well understood. 

For the final selection, we define search regions around the known \bs\
and \bd\ masses~\cite{PDG2010}. These regions correspond to approximately 
$\pm 2.5\sigma_{m}$, where $\sigma_{m}\approx24$~$\mathrm{MeV}/c^{2}$ 
is the estimated two-track  mass resolution.
The sideband regions $5.0 < \Mmm < 5.169\:\GeV/c^2$ and  $5.469 < \Mmm
< 5.969\:\GeV/c^2$ are used to estimate combinatorial backgrounds.
Backgrounds from $B \ra \hh$ decays (where $h$, $h^{\prime} = \pi^{\pm}$ or $K^\pm$), 
which peak in the signal mass region, are estimated separately.

Fourteen variables are used to construct a NN
discriminant \nn\ that ranges from 0 to 1 and enhances the signal-to-background ratio~\cite{NeuroBayes}.
The variables include dimuon vertex related information (e.g.  $\lambda/\sigma_\lambda$), 
the impact parameters with respect to the primary vertex and 
transverse momenta of the muons, the isolation of the {\it B} candidate, and the opening angle $\pting$.
The NN is trained with background events sampled from the sideband regions and signal events
generated with a simulation described below.
Only a fraction of the total number of background and simulated signal
events are used to train the NN.  The remainder are used to test for NN overtraining and
to determine the signal and background efficiencies.
Several tests are done to ensure \nn\ is independent of \Mmm.

All selection criteria were finalized before revealing the content of the signal regions.  
The optimization of the criteria used the expected upper limit on the \bsmm\ branching fraction as a figure of merit.
To exploit the difference in the \Mmm\ distributions between signal and background 
and the improved suppression of combinatorial background at large \nn, the data is divided into 
sub-samples in the $\left(\nn,\Mmm\right)$ plane.
The CC and CF samples are each divided into 40 sub-samples.  There are eight bins in \nn\
with bin boundaries $0.70, 0.76, 0.85, 0.90, 0.94, 0.97, 0.987, 0.995$ and $1$.
Within each \nn\ bin we employ five \Mmm\ bins, each 24~MeV/$c^2$ wide, centered on the world 
average \bsd\ mass.  
The expected backgrounds and efficiencies are calculated in each bin separately.

For measuring efficiencies, estimating backgrounds, and optimizing the
analysis, samples of $\bsdmm$, $\bjk$, and $B \ra \hh$ are generated with the {\sc pythia} 
program~\cite{pythia} and a \cdf\ detector simulation.  The $p_{T}$ 
spectrum and the \iso\ distribution of the $B$-mesons are weighted to match distributions 
measured in samples of \bjk\ and \bsjp\ events~\cite{Acosta:2004yw}.

We use a relative normalization to determine the \bsmm\ branching
fraction:
\begin{eqnarray}\label{eq:brbsmm}
\brbsmm = \frac{N_s}{N_{+}}\:
     \frac{\alpha_{+}}{\alpha_{s}}
     \:\frac{\epsilon_{+}}{\epsilon_{s}} \:
     \frac{1}{\epsilon_{N}} \: \nonumber
     \frac{f_{+}}{f_{s}}  
     \: \mathcal{B}(B^+),
\end{eqnarray}
where  $N_s$ is the number of \bsmm\ candidate events.
The observed number of \bjk\ candidates is $N_{+} = 22\,388\pm 196$ and $9\,943\pm 138$ in 
the CC and CF channels, respectively. The contribution of \bjpi\ events is negligible.
We use $\mathcal{B}(B^+) = \mathcal{B}(\bjk\ra\mm K^{+})=(6.01\pm0.21)\times 10^{-5}$~\cite{PDG2010} 
and the ratio of $B$-meson production fractions ${f_{+}}/{f_{s}}=3.55\pm0.47$~\cite{PDG2010}. 
The parameter $\alpha_{s}$ ($\alpha_{+}$) is the acceptance of the trigger and  
$\epsilon_{s}$ ($\epsilon_{+}$) is the efficiency of the
reconstruction requirements for the signal (normalization) mode.
The reconstruction efficiency
includes trigger, track, muon, and baseline requirement efficiencies.  
The NN efficiency ${\epsilon_{N}}$ only applies to the signal mode
since it is not used to select the \bjk\ sample.
The expression for \brbdmm\ is derived by replacing \bs\ with \bd\ 
and ${f_{+}}/{f_{s}}$ with ${f_{+}}/{f_{d}}=1$.
The ratios of acceptances ${\alpha_{+}}/{\alpha_{s}}$ are 
$0.307\pm0.018$ and $0.197\pm0.014$ for the CC and CF topologies, respectively.
These ratios are measured using simulated events. The uncertainties include 
contributions from systematic variations
of the modeling of the $B$-meson $p_{T}$ distributions and the longitudinal 
beam profile.
The ratio of reconstruction efficiencies is
${\epsilon_{+}}/{\epsilon_{s}}=0.81\pm0.03$ as determined from studies
using samples of $J/\psi\ra\mm$ and \bjk\ events collected with the same triggers. 
The uncertainty in ${\epsilon_{+}}/{\epsilon_{s}}$ is dominated by kinematic differences between 
$J/\psi\ra\mm$ and \bsdmm\ decays.  The $\epsilon_{N}$ is estimated from the simulation.  
We assign a relative systematic uncertainty on $\epsilon_{N}$ of 4-7\% , depending on \nn\ bin, 
using comparisons of the NN performance in 
simulated and observed $\bjk$ event samples, and the 
statistical uncertainty on studies of the $p_T$ and $\iso$ distributions from observed 
$\bsjp$ event samples.  The \bdmm\ decay is determined to have the same acceptances and 
efficiencies. Treating CC and CF together, about 90\% of simulated \bsmm\ events surviving the initial 
requirements have $\nn > 0.70$, with about 45\% having $\nn > 0.995$.  The expected SM yield of 
\bsmm\ events ranges from $0.05$ in the lowest \nn\ bin to $1.0$ events in the 
highest \nn\ bin summing the CC and CF contributions.  The expected SM yield of \bdmm\ events 
is about thirty times smaller.

\begin{figure*}[th]
  \centering
  \vspace*{-0.0in}
  \includegraphics[width=3.0in]{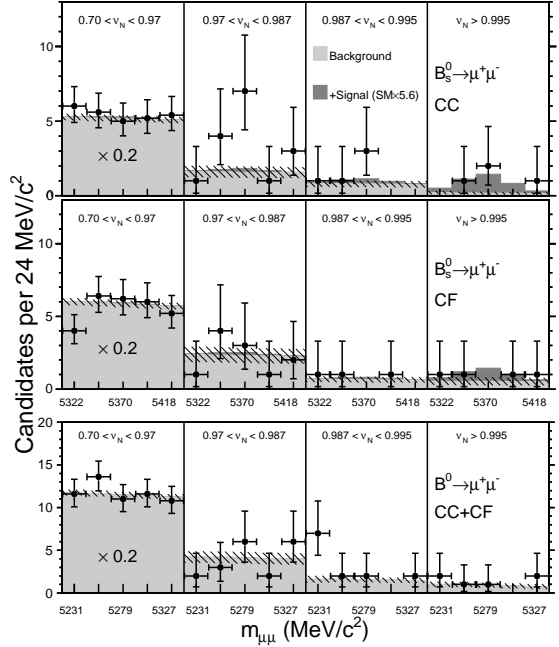}
  \caption{\label{fig:Bs} For the \bs\ and \bd\ signal regions, the observed number of 
    events (points) is compared to the total expected background (light
    grey) and its uncertainty (hatched) using the $\left(\nn , \Mmm\right)$ bins from the optimization.  The background
    uncertainty is the quadrature sum of the relevant systematic uncertainties. The top and middle rows show the results in the 
    \bs\ mass signal region for the CC and CF channels, respectively.  The bottom row 
    shows the results in the \bd\ mass signal region for the CC and CF channels combined. 
    The results for the first 5 \nn\ bins are combined (and scaled by $0.2$) while the results
    for the last three bins are each shown separately. Also shown is the expected contribution
    from \bsmm\ events (dark gray) using a branching fraction that corresponds to the 
    central value from the fit to the data, which is 5.6 times the expected SM value.
  }
\end{figure*}

The expected background is obtained by summing contributions from
the combinatorial background and from $B\ra\hh$ decays.
To estimate the combinatorial background, we fit the \Mmm\ distribution of sideband events with
$\nn > 0.70$ to a linear function.
We only use events with \Mmm $>$ 5~GeV/$c^2$ in order to suppress
contributions from $b\rightarrow \mu^+\mu^- X$ decays.
The slopes are then fixed, and the normalization is
determined for each \nn\ bin separately using the relevant sideband
events.  In addition to the statistical uncertainties of the slope and normalization parameters,
systematic uncertainties are assigned by comparing results derived using alternative fit functions 
and ranges.  The systematic uncertainties vary from about 7\% for the lower \nn\ bins 
to about 45\% for the highest \nn\ bins. 
The $B\ra\hh$ contributions are estimated using efficiencies
determined from the simulation,  probabilities of misidentifying hadrons as
muons measured in data, and normalizations derived from their branching
fractions~\cite{bhhprl,PDG2010}.  The hadron misidentification probabilities are
parameterized as a function of hadron $p_{T}$ and instantaneous luminosity using
a $D^0 \rightarrow K^- \pi^+$ data sample obtained from
$D^{*+}\rightarrow D^0\pi^+$ decays.  In addition to the statistical uncertainties 
from the $D^{0}$ sample, systematic uncertainties are assigned to account for 
residual variations of the misidentification probability due to variations in detector performance (primarily arising from occupancy and calibration effects) and for
branching fraction uncertainties.  For the \bs\ modes there is an additional uncertainty from 
$f_+ / f_s$.
The estimated  $B\ra\hh$ background is approximately one quarter of the
total background in the \bdmm\ search while in the \bsmm\ search it is 
a factor of ten smaller than both the combinatorial background and the SM signal.
The expected background is shown in Fig.~\ref{fig:Bs} for the \bsmm\ 
and \bdmm\ searches.  
The background estimates are cross-checked using
three sets of independent control samples:  \mm\ events with $\lambda<0$ and $\mu^{\pm}\mu^{\pm}$ events,
both of which are dominated by combinatorial backgrounds,  and
a misidentified-muon enhanced \mm\ sample with at least one muon 
candidate failing the muon quality requirements.  The latter sample has a significant 
contribution from $B\ra\hh$ backgrounds.  
We compare the predicted and observed
number of events in each of these control samples for all 80 sub-samples and observe no significant 
discrepancies.

Two fits are performed on the data, a background-only fit $(b)$ and a signal-plus-background 
fit $(s+b)$ for which the branching fraction of the signal is left floating. 
A log-likelihood ratio is formed, $-2\ln Q$,
where $Q = \mathcal{L}(s+b | {\rm{data}})/\mathcal{L}(b | {\rm{data}})$ and $\mathcal{L}(h|x)$ is
the likelihood of hypothesis $h$ given observation $x$;  this likelihood is obtained by
multiplying Poisson probabilities over all 80 sub-samples and is minimized with respect to the
nuisance parameters that model our systematic uncertainties.
To evaluate the consistency of the data in the
signal region with our background model, we compare the observed value of $-2\ln Q$ with the
distribution of $-2\ln Q$ obtained from an ensemble of background-only simulated experiments.
The effects of systematic uncertainties are included in the
simulated experiments by randomly choosing the nuisance parameters from Gaussian distributions.  
The fraction of simulated experiments with a value of $-2\ln Q$ less than that observed in the 
data is used to determine the $p$-value for the background-only hypothesis.  

The data in the signal regions are shown in Fig.~\ref{fig:Bs} using the 
$\left(\nn , \Mmm \right)$ binning from the optimization.  In the \bd\ search region the data are consistent with the background 
prediction and have a $p$-value of 23\%.  In the \bs\ search region the data exceed the 
background prediction and have a $p$-value of $0.27\%$.  The excess is concentrated in bins 
with $\nn > 0.97$.  If we restrict ourselves to only the two highest \nn\ bins ($\nn > 0.987$), 
which together account for 85\% of the signal acceptance, we find a $p$-value
of 0.66\%.  For the \bsmm\ analysis we also produce an ensemble of simulated experiments that 
includes a \bsmm\ contribution at the expected SM branching fraction~\cite{smbr} and yields a 
$p$-value of $1.9\%$.  The corresponding $p$-value for the two highest \nn\ bins alone is 4.3\%.

We use a modified frequentist approach~\cite{TRJNim, ALRead} that includes the effects of 
systematic uncertainties to calculate expected and observed limits.
We calculate expected limits of $\brbdmm < 4.6\times 10^{-9}$ and $\brbsmm < 1.5\times 10^{-8}$ 
at the 95\% confidence level (C.L.), a factor 3.3 improvement relative to
our previous analysis~\cite{CDFPRL2008}. We calculate observed limits of 
$\brbdmm < 6.0 (5.0) \times 10^{-9}$ and $\brbsmm < 4.0 (3.5) \times 10^{-8}$ at 
95\% (90\%) C.L.   If we assume the observed excess in the \bs\ region is due to signal, 
we determine \brbsmm=$(1.8^{+1.1}_{-0.9})\times 10^{-8}$ using the data $-2\ln Q$ distribution
and taking the central value from the minimum and the associated uncertainty as the interval 
corresponding to a change of one unit.  By examining the interval corresponding to a change
of $2.71$ units we set bounds of $ 4.6 \times 10^{-9} < \brbsmm < 3.9 \times 10^{-8}$ at the
90\% C.L.  As a cross check we use a Bayesian technique to make a point estimate and to derive
bounds at 90\% C.L. and obtain results very similar to those reported here.  Using the central value
for the fitted \bsmm\ branching fraction we produce an ensemble of simulated experiments and find 
a $p$-value of $50\%$. 

The source of the data excess in the $0.970 < \nn <0.987$ bin of the \bs\ signal region is investigated. The same events, same fits, and same methodologies are used for both the \bs\ and \bd\ searches.  Because the data in the \bd\ search region shows no excess, problems with the background estimates are ruled out.  In particular,
the only peaking background in this mass region is from $B\ra\hh$ decays, whose contribution
to the \bd\ search region is ten times larger than to the \bs\ search region.  Problems with
the NN are ruled out by the many studies performed.  These NN studies find no evidence of a
\nn - \Mmm\ correlation, no evidence of overtraining, and no evidence of a significant 
mis-modeling of the \nn\ shape, even in the region $0.995<\nn$.  In short, there is no evidence
that the excess in this bin is caused by a mistake or systematic error in our background 
estimates or our modeling of the \nn\ performance and distribution.  The most plausible remaining
explanation is that this is a statistical fluctuation.  For our central result we use 
the full set of bins that had been established a priori since this represents an unbiased choice.
As discussed above, if we remove the $0.970 < \nn <0.987$ bin the results
are not significantly affected.

In summary, we have performed a search for \bdmm\ and \bsmm\ decays 
using 7~fb$^{-1}$ of integrated luminosity collected by the \cdf\ detector at the Fermilab
Tevatron. The data in the \bd\ search region are consistent with background 
expectations and the world's most stringent upper limit on \brbdmm\ is established.
The data in the \bs\ search region are in excess of the background predictions with a
$p$-value of $0.27\%$.  
A fit to the data determines  \brbsmm $=(1.8^{+1.1}_{-0.9})\times 10^{-8}$ including all
uncertainties.  

\section{\bf Acknowledgments}
We thank the Fermilab staff and the technical staffs of the participating 
institutions for their vital contributions. This work was supported by 
the U.S. Department of Energy and National Science Foundation; the 
Italian Istituto Nazionale di Fisica Nucleare; the Ministry of Education, 
Culture, Sports, Science and Technology of Japan; the Natural Sciences 
and Engineering Research Council of Canada; the National Science Council 
of the Republic of China; the Swiss National Science Foundation; the A.P. 
Sloan Foundation; the Bundesministerium f\"ur Bildung und Forschung, 
Germany; the Korean World Class University Program, the National Research Foundation
of Korea; the Science and Technology Facilities Council and the Royal 
Society, UK; the Institut National de Physique Nucleaire et 
Physique des Particules/CNRS; the Russian Foundation for Basic Research; the
Ministerio de Ciencia e Innovaci\'{o}n, and Programa Consolider-Ingenio 2010, Spain;
the Slovak R\&D Agency; the Academy of Finland; and the Australian Research
Council (ARC).



\begin{thebibliography}{99.}

\bibitem{qconj}
Throughout this paper inclusion of charge conjugate modes is implied.

\bibitem{smbr}
E.~Gamiz, {\it{et al.}} (HPQCD Collaboration), 
Phys.\ Rev.\ D {\bf{80}}, 014503 (2009);
A.~J.~Buras, M.~V.~Carlucci, S.~Gori, and G.~Isidori, 
J. High Energy Phys. 10, 009 (2010).

\bibitem{tanb}
C.~Hamzaoui, M.~Pospelov, M.~Toharia, Phys.\ Rev.\ D {\bf{59}}, 095005 (1999);
 S.~R.~Choudhury and N.~Gaur, Phys.\ Lett.\ B {\bf{451}}, 86 (1999);
 K.S.~Babu and C.~Kolda, Phys.\ Rev.\ Lett.\ {\bf{84}}, 228 (2000).

\bibitem{CDFPRL2008}
  T.~Aaltonen {\it{et al.}} (CDF~Collaboration),
  Phys.\ Rev.\ Lett.\ {\bf{100}}, 101802 (2008).

\bibitem{D0PLB2010}
 V.~Abazov {\it{et al.}} (D0 Collaboration), 
 Phys.\ Lett.\ B {\bf{693}}, 539 (2010).

\bibitem{LHCb}
 R.~Aaij {\it{et al.}} (LHCb Collaboration), Phys.\ Lett.\ B {\bf{699}}, 330 (2011).

\bibitem{rparity}
 A.~Dedes, H. K.~Dreiner, and U.~Nierste, Phys.\ Rev.\ Lett.\ {\bf{87}}, 251804 (2001);
 R.~Arnowitt {\it{et al.}}, Phys.\ Lett.\ B {\bf{538}}, 121 (2002).

\bibitem{dark}
 S.~Baek {\it{et al.}}, J. High Energy Phys. 06, 017 (2005).

\bibitem{chi2} 
 R.~Ruiz de Austri, R.~Trotta and L.~Roszkowski, J. High Energy Phys. 05, 002 (2006);
 J.~Ellis {\it{et al.}}, J. High Energy Phys. 05, 063 (2006).

\bibitem{baek}
S.~Baek, P.~Ko and W.Y.~Song, Phys.\ Rev.\  Lett.\ {\bf 89}, 271801 (2002).

\bibitem{buras}
A.~Buras {\it{et al.}}, J.\ Phys.\ Conf.\ Ser.\ {\bf 171}, 012004 (2009).

\bibitem{Acosta:2004yw}
  D.~Acosta {\it et al.} (CDF~Collaboration),
  Phys.\ Rev.\ D {\bf 71}, 032001 (2005).

\bibitem{PRL1}
  D.~Acosta {\it{et al.}} (CDF~Collaboration), 
  Phys.\ Rev.\ Lett.\ {\bf{93}}, 032001 (2004).

\bibitem{mixprl}
 A.~Abulencia {\it et al.} (CDF~Collaboration),
 Phys.\ Rev.\ Lett.\ {\bf{97}}, 242003 (2006).

\bibitem{bhhprl}
  T.~Aaltonen {\it et al.}  (CDF~Collaboration),
  Phys.\ Rev.\ Lett.\ {\bf{106}}, 181802 (2011).

\bibitem{isodef} $\iso=| \ptmm | /(\sum_{i}p^{i}_{T} + | \ptmm | )$, where
 \pmm\ is the momentum of the dimuon pair; the
 sum is over all tracks with $\sqrt{(\Delta\eta)^2+(\Delta\phi)^2} \le 1$;
 $\Delta\phi$ and $\Delta\eta$ are the relative azimuthal angle and
 pseudorapidity of track $i$ with respect to \pmm.
  
\bibitem{PDG2010}
  K.~Nakamura {\it et al.} (Particle Data Group),
  J.\ Phys.\ G {\bf 37}, 075021 (2010).

\bibitem{NeuroBayes}
 M.\ Feindt and U.\ Kerzel, Nucl.\ Instrum.\ Methods\ A {\bf{559}}, 190
	(2006).

\bibitem{pythia}
 T.~Sj\"{o}strand {\it{et al.}}, Comp.\ Phys.\ Comm.\ {\bf{135}}, 238 (2001).

\bibitem{TRJNim} 
 T.~Junk, Nucl.\ Instrum.\ Methods\ A {\bf{434}}, 435 (1999).

\bibitem{ALRead}
 A.L.\ Read, ``Modified frequentist analysis of search results (the $CL_s$ method)'', 
 in F.\ James, L.\ Lyons and Y.\ Perrin (eds.), {\it{Workshop on Confidence Limits}}, 
 CERN Yellow Report 2000-005 (2000).


\end{thebibliography}

\end{document}